\documentclass[epj]{svjour}
\usepackage{graphics}

\begin{document}
\title {Electron motion in a Holstein molecular chain in an electric field}

\author{V.D. Lakhno\inst{1}\thanks{\emph e-mail: lak@impb.psn.ru}
\and A.N. Korshunova\inst{1}\thanks{\emph e-mail: alya@impb.psn.ru}}

\institute{%
Institute of Mathematical Problems of Biology, Russian Academy of
Sciences,  Pushchino, Moscow Region, 142290, Russia}%

%

\abstract{
A charge motion in an electric field in a Holstein molecular chain is modeled in the absence of dissipation.  It is shown that in a weak electric field a Holstein polaron moves uniformly experiencing small oscillations of its shape. These oscillations are associated with the chain's discreteness and caused by the presence of Peierls-Nabarro potential there. The critical value of the electric field intensity at which the moving polaron starts oscillating at Bloch frequency is found. It is shown that the polaron can demonstrate Bloch oscillations retaining its shape. It is also shown that a breathing mode of Bloch oscillations can arise. In all cases the polaron motion along the chain is infinite.
} 

\maketitle

Presently, there are a lot of papers devoted to modeling the motion of a charged particle in various-type molecular chains (see books and reviews \cite{1}-\cite{5} and references therein). Interest in the problem is associated with possible practical applications of such chains in nanoelectronics \cite{6},\cite{7}. Functioning of nanoelectronic devices is based on the motion of charges along molecular chains placed in an external electric field. Notwithstanding the great number of publications on the problem of charge motion in a molecular chain, it would be an exaggeration to say that the problem is conclusively solved.

It is well known that an electron in an ideal non-deformable molecular chain placed in an electric field demonstrates Bloch oscillations \cite{8}-\cite{11}. In a deformable molecular chain an electron turns into a polaron state. In the simplest case of a Holstein molecular chain in the absence of an external field the motion of an electron with steady velocity $v$ appears to be impossible. The reason is that a steadily moving polaron state in a crystal with dispersion-free phonons is not the eigenstate of the Holstein Hamiltonian \cite{12}. Using a discrete Holstein molecular chain as an example, in \cite{13} we showed that in a weak electric field a polaron as a whole demonstrates Bloch oscillations which in a strong electric field take the form of breathing mode of Bloch oscillations. In contrast to this, in \cite{12} it was shown that in a continuum approximation in an electric field not exceeding a certain critical value, the polaron motion can be steady.

The aim of this paper is to model numerically the charge steady motion in a discrete chain in an electric field and to study the conditions under which it turns into oscillatory motion.
In the Holstein model under consideration the electron motion in a molecular chain is described by Hamiltonian \cite{14}-\cite{18}:
\begin{eqnarray}\label{1}
    \nonumber&\hat{\mathrm{H}}=\nu\sum\nolimits_{n}\bigl(a_n^+a_{n-1}+a_n^+a_{n+1}\bigr)
    +\alpha'\sum\nolimits_{n}q_na_n^+a_n\\
    &+\sum\nolimits_{n}e\mathcal{E}ana_n^+a_n +\sum\nolimits_{n}\hat{\mathrm{P}}_n^2\big/2M+\sum\nolimits_{n}Kq_n^2\big/2,
\end{eqnarray}
where $\nu$ is the matrix element of the charge transition from the $n$-th site to the $n\pm1$ site, $a_n^+,a_n$ are the operators of the birth and annihilation of the charge at a site with number $n$, $\alpha'$ is a constant of the charge interaction with site displacements $q_n$, $e$ is the electron charge, $\mathcal{E}$ is the electric field intensity, $a$ is the lattice constant, $\hat{\mathrm{P}}_n=-i\hbar\, \partial\big/\partial q_n$, $M$ is the site's effective mass, $K$ is the elastic constant.

To study the dynamics of the charge motion in the chain use is usually made of a semiclassical description based on the fact that the site's mass $M$ exceeds greatly the electron mass. In order to pass on to semiclassical description we will seek the wave function of the system $|\Psi\rangle$ in the form of its expansion in terms of coherent states:
\begin{equation}\label{2}
    |\Psi(t)\rangle\!\!=\!\!\!\sum_{n}\!b_n(t)a_n^+\exp\Bigl\{\!-\frac{i}{\hbar}\!\sum_{j}
    \bigl[\beta_j(t)\hat{\mathrm{P}}_j\!-\!\pi(t)q_j\bigr]\!\Bigr\}|0\rangle\!,
\end{equation}
where $|0\rangle$ is the vacuum wave function, while the quantities $\beta_j(t)$ and $\pi_j(t)$ satisfy the relations:
\begin{equation}\label{3}
    \langle\Psi(t)|q_n|\Psi(t)\rangle=\beta_n(t),\ \ \langle\Psi(t)|\hat{\mathrm{P}}_n|\Psi(t)\rangle=\pi_n(t).
\end{equation}
Dynamical equations for the quantities $b_n(t)$ and $\beta_n(t)$ obtained from (\ref{1})-(\ref{3}) have the form:
\begin{eqnarray}\label{4}
    i\hbar \dot{b}_n&=&\nu(b_{n-1}+b_{n+1})+\alpha'\beta_nb_n+e\mathcal{E}anb_n,\\
    M\ddot{\beta}_n&=&-K\beta_n-\alpha'|b_n|^2.\label{5}
\end{eqnarray}
Equations (\ref{4}) are Schr\"{o}dinger equations where $b_n$ are amplitudes of the probabilities of the charge localization at the $n$-th site; (\ref{5}) are classical motion equations for sites displacements $\beta_n$.

Turning to coherent states using transformation (\ref{2}) corresponds to the idea of the strong coupling polaron. Herewith the quantities $\beta_n$ determine the phonon "environment" surrounding the particle. Notice that in the absence of an electric field, the steady motion of the polaron along the chain is impossible. The reason is that the group velocity of phonons $v_g=\partial \Omega\big/\partial k$ is equal to zero since the quantity $\Omega$, according to (\ref{5}), does not depend on the wave vector $k$: $\Omega=\sqrt{K\big/M}$. In other words, in the absence of dispersion in the chain, the polaron "environment" cannot follow the particle's motion along the chain.

\begin{figure}[t]
\resizebox{0.43\textwidth}{!}{\includegraphics{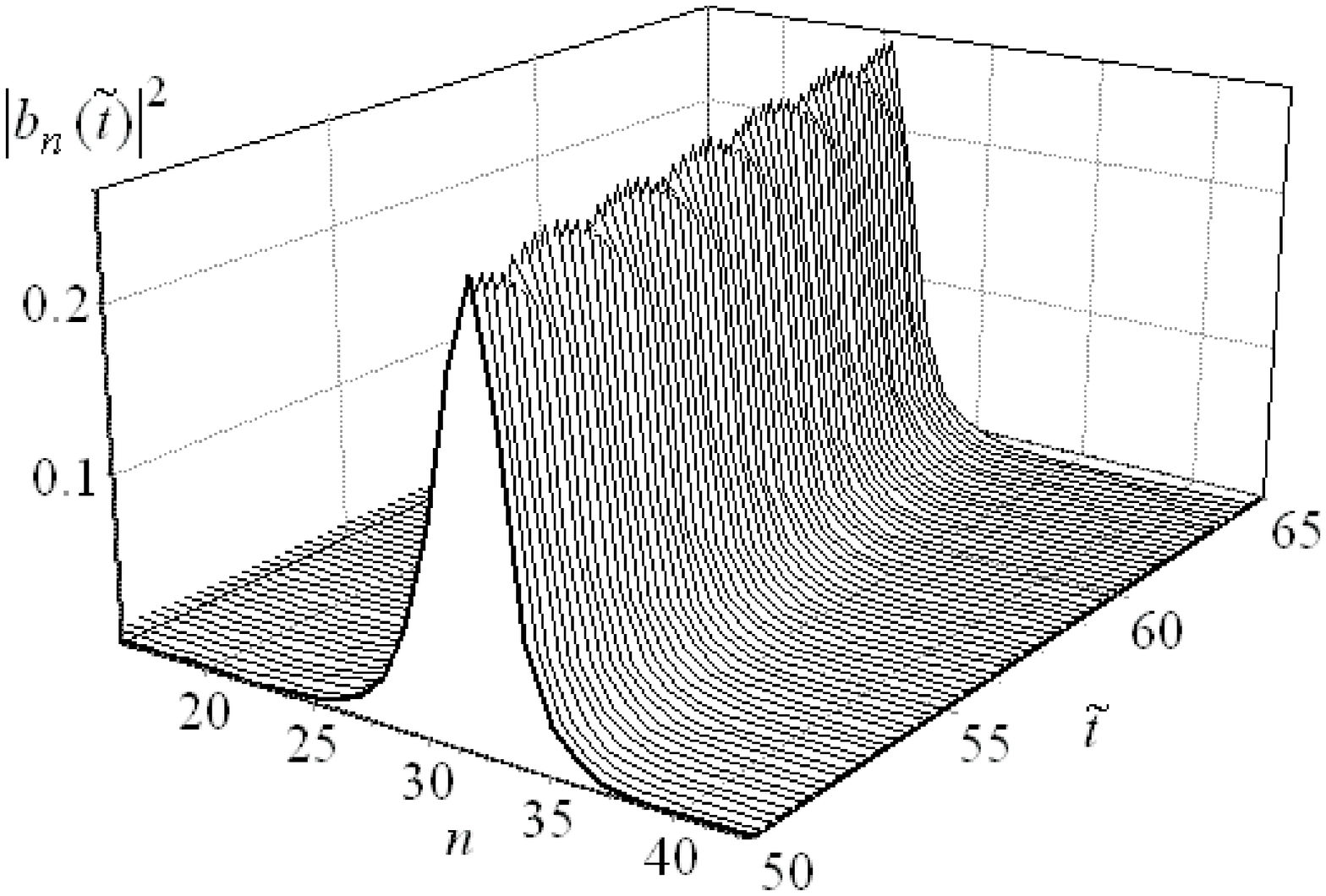}}(a)\\
\resizebox{0.43\textwidth}{!}{\includegraphics{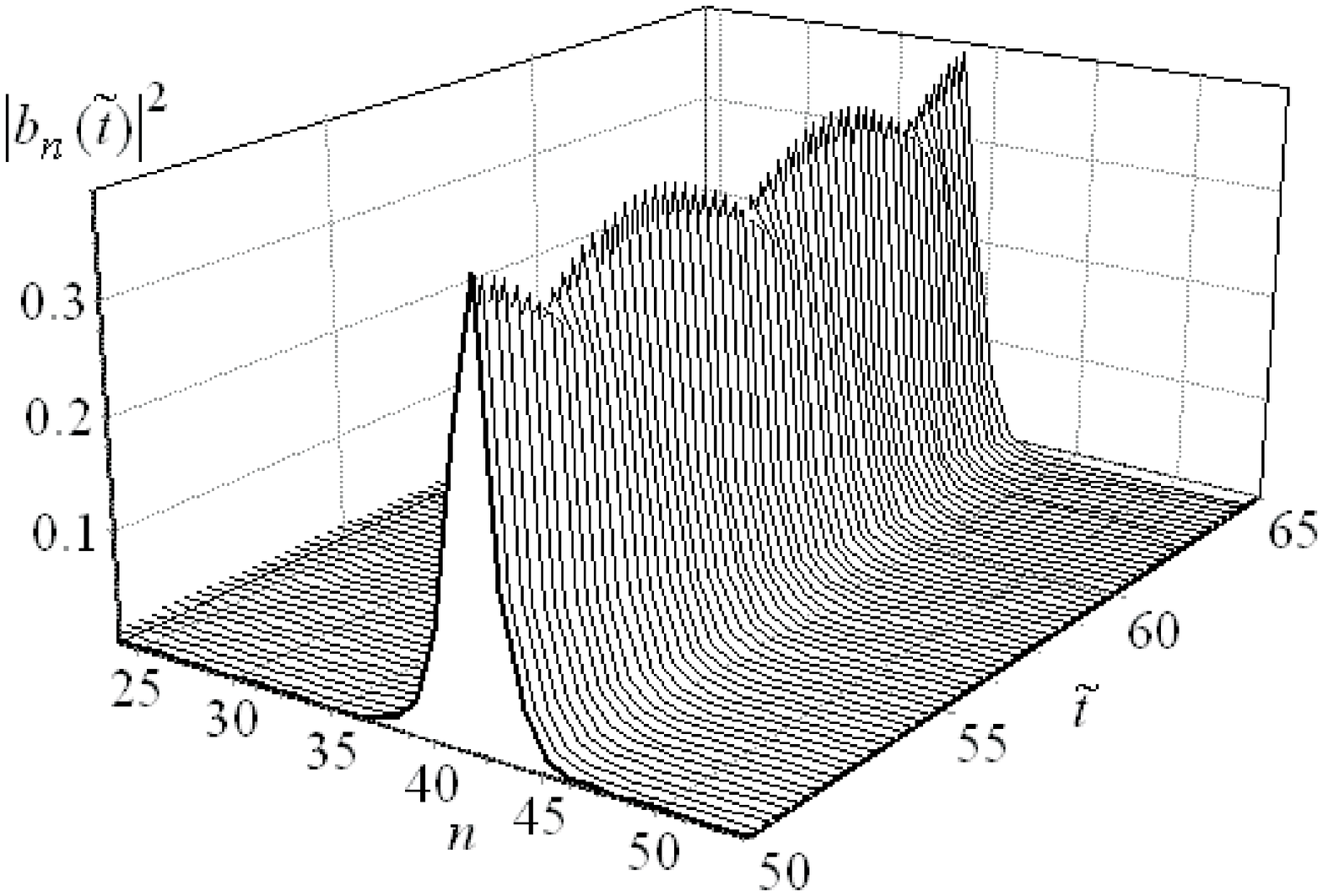}}(b)
\caption{Motion of the polaron at a constant velocity in an electric field of intensity $E=0.01$ at $\omega=1, \eta=1.276$ for various values of the parameter $\kappa$: (a)~$\kappa=2$, (b)~$\kappa=3$. Oscillations of the polaron shape whose amplitude depends on the values of $\kappa$ are conspicuous (here and elsewhere $\tau=10^{-14}$sec.)
}\label{fig_1}
\end{figure}
For numerical modeling of the polaron motion, we used dimensionless variables in which system (\ref{4}),(\ref{5}) has the form:
\begin{eqnarray}\label{6}
    i\frac{db_n}{d\tilde{t}}&=&\eta\bigl(b_{n+1}+b_{n-1}\bigr)+\kappa\omega^2u_nb_n+Enb_n,\\
    \frac{d^{\,2}u_n}{d\tilde{t}^2}&=&-\omega^2u_n-|b_n|^2,\label{7}
\end{eqnarray}
where dimensional quantities involved in (\ref{4}),(\ref{5}) are related to dimensionless quantities involved in (\ref{6}),(\ref{7}) as:
\begin{eqnarray}\label{8}
    \eta&=&\tau\nu/\hbar,\ \omega^2=\tau^2K\big/M,\ q_n=\beta u_n,\ t=\tau\tilde{t},\ \\
    \nonumber \kappa\omega^2&=&\tau^3(\alpha')^2\big/M\hbar,\ \beta=\tau^2\alpha'\big/M,\ E=\mathcal{E}ea\tau\big/\hbar,
\end{eqnarray}
where $\tau$ is an arbitrary time scale. In view of arbitrariness of $\tau$ any set of dimensionless parameters $\eta, \omega, \kappa, E$ correspond to a continuum of possible dimensional parameters of the molecular chain: $\nu, K, M, \alpha', \mathcal{E}$.

\begin{figure}[h]
\begin{center}
\resizebox{0.4\textwidth}{!}{\includegraphics{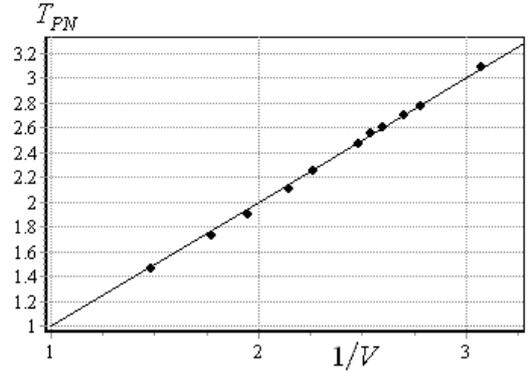}}
\caption{Numerical $(\bullet)$ comparison of the oscillation period
$T_{PN}$ with the quantity~$1/V$ for $E = $ $0.02$, $0.015$, $0.01$,
$0.005$, $0.003$, $0.001$, $0.0007$, $0.0005$, $0.0003$, $0.0002$,
$0.0001$ at the following parameter values: $\kappa=2$, $\omega=1,
\eta=1.276.$ The right top point $(\bullet)$ on the graph
corresponds to the value $E=0.0001$. The straight line corresponds
to the graph $T_{PN}=1/V$. }\label{fig_2}
\end{center}
\end{figure}
In the continuum limit, solution of (\ref{6}),(\ref{7}) yields the following relation between the polaron equilibrium velocity $V=va/\tau$ and the electric field intensity \cite{12}:
\begin{equation}\label{9}
    E=2\pi^2\frac{\omega^4\eta^2}{\kappa}\frac{1}{V^4}\frac{1}{\sinh^2(2\pi\eta\,\omega/\kappa
    V)}.
\end{equation}
According to \cite{12}, the steady motion of a polaron in the conservative system under consideration in an electric field appears possible due to the fact that while the polaron moves along the chain it leaves behind a "tail" of excited oscillators. In this process the work done by the electric field is expended for excitation of these oscillators.

Figure~\ref{fig_1} shows the polaron motion in an electric field for various values of the parameter $\kappa$, obtained as a result of numerical solution of discrete equations (\ref{6}),(\ref{7}).

The calculations were carried out by a standard 4-th order Runge-Kutta numerical method. The initial values of $|\,b_n(0)|$ $(b_n=x_n+iy_n)$ were chosen in the form of an inverse hyperbolic cosine:
\begin{eqnarray}\label{10}
 &|\,b_n(0)|=\frac{\sqrt{2}}{4}\sqrt{\frac{\kappa}{|\,\eta|}}\,\mathrm{ch}^{-1}
 \Bigl(\frac{\kappa(n-n_0)}{4|\,\eta|}\Bigr), \\ \nonumber
 x_n^0&=|\,b_n(0)|(-1)^n\big/\sqrt{2},\ \ y_n^0=|\,b_n(0)|(-1)^{n+1}\big/\sqrt{2},
\end{eqnarray}
$u_n^0=|\,b_n(0)|^2\big/\omega_n^2$, $du_n^0\big/d\tilde{t}=0$.
The parameter values were taken to be: $\omega=1, \eta=1.276$.
The chains length was fitted to suit the calculations duration with the proviso that the polaron should not come too close to the end of a chain when calculations are over. To calculate the numerical dependence of the polaron velocity $V$ on the electric field intensity $E$ the sequence length was taken in the range from $N=5000$ to $N=30000$ depending on $E$. The value of $n_0$ (the center of an initial inverse hyperbolic cosine) in (\ref{10}) was fitted so that at the beginning of the calculations the polaron be far away from the end of the chain. For the parameter values  chosen in this work it was appropriate to take $n_0$ from $n_0=300$ to $n_0=700$.

To calcalculate numerically the polaron velocity $V$ we used the relations:
\begin{equation}\label{11}
 V=dX\big/d\tilde{t},\;\; X(\tilde{t})=\sum\nolimits_{n}n|b_n(\tilde{t})|^2.
\end{equation}

Figure~\ref{fig_1} suggests that while moving steadily, the polaron shape oscillates periodically every  $T_{PN}=1\big /V$ due to the presence of Peierls-Nabarro potential in a discrete chain. The amplitude of these oscillations turns to zero in the continuum  approximation when the polaron size exceeds considerably the distance between neighboring sites. Numerical comparison of the oscillation period $T_{PN}(E)$ with the quantity $1\big/V(E)$ for various values of the electric field intensity $E$ is given in Figure~\ref{fig_2}.

\begin{figure}[h]
\resizebox{0.4\textwidth}{!}{\includegraphics{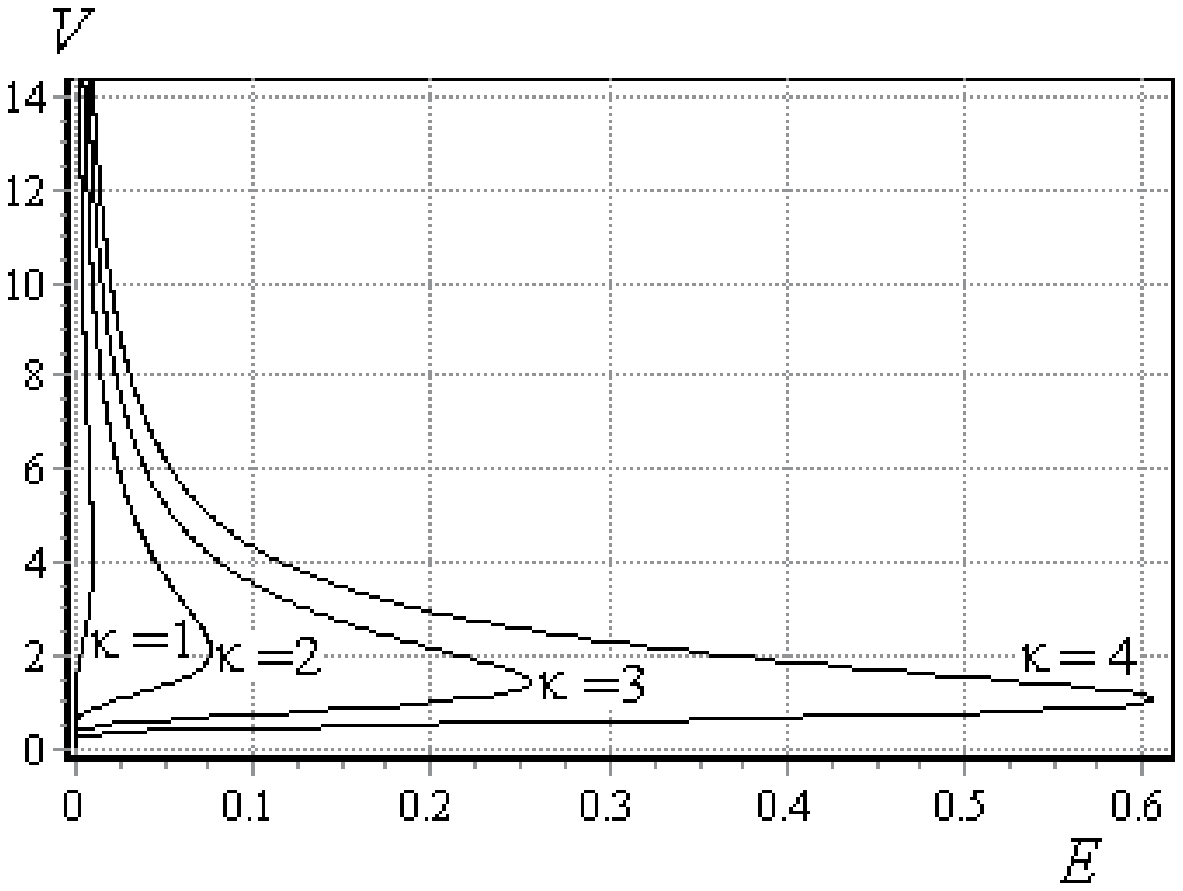}}(a)\\
\resizebox{0.4\textwidth}{!}{\includegraphics{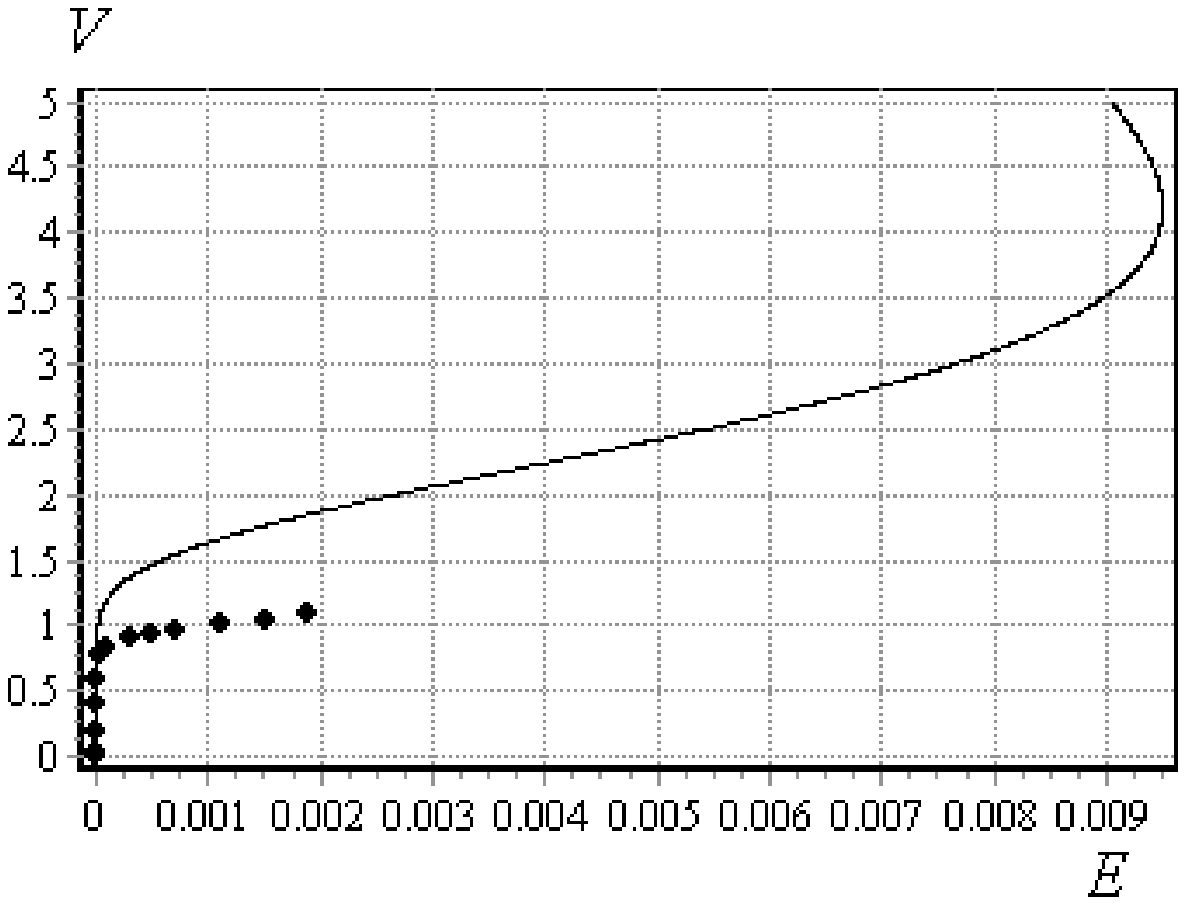}}(b)
\caption{(a) - Dependence (\ref{9}) of the polaron velocity $V$ on the electric field intensity $E$ for various values of the parameter $\kappa$ ($\kappa=1,2,3,4), \eta=1.276, \omega=1$.
(b) - Comparison of the theoretical (---) (\ref{9}) and numerical $(\bullet)$ (\ref{6}),(\ref{7}) dependence of the polaron velocity $V$ on the electric field intensity $E$ for $\kappa=1, \eta=1.276, \omega=1$.
}\label{fig_3}
\end{figure}

Figure~\ref{fig_3}a demonstrates the dependencies of the polaron velocity $V$ on the field intensity $E$ determined by relation (\ref{9}) for various values of the parameter $\kappa$. All the $V(E)$ curves have an infinite derivative at zero point: $V'(0)=\infty$ which corresponds to infinite mobility of the polaron in the conservative system under consideration determined by Hamiltonian (\ref{1}). Figure~\ref{fig_3}b compares the average velocity of the polaron steady motion obtained by numerical integration of system (\ref{6}),(\ref{7}) with the theoretical curve determined by (\ref{9}) for the following parameter values: $\kappa=1, \eta=1.276, \omega=1$. For these values of the parameters $\kappa$ and $\eta$, continuum approximation is fulfilled to a high accuracy since the polaron state radius: $\tilde{r}=r/a=4\eta/\kappa\approx5.1\gg1$. It is seen from Figure~\ref{fig_3}b that the difference between the theoretical and experimental values increases as $V$ grows. This difference is likely to be due to the fact that in deriving the theoretical dependence (\ref{9}) we did not take into account that the polaron shape changes as velocity $V$ increases.

\begin{figure}[t]
\begin{center}
\resizebox{0.43\textwidth}{!}{\includegraphics{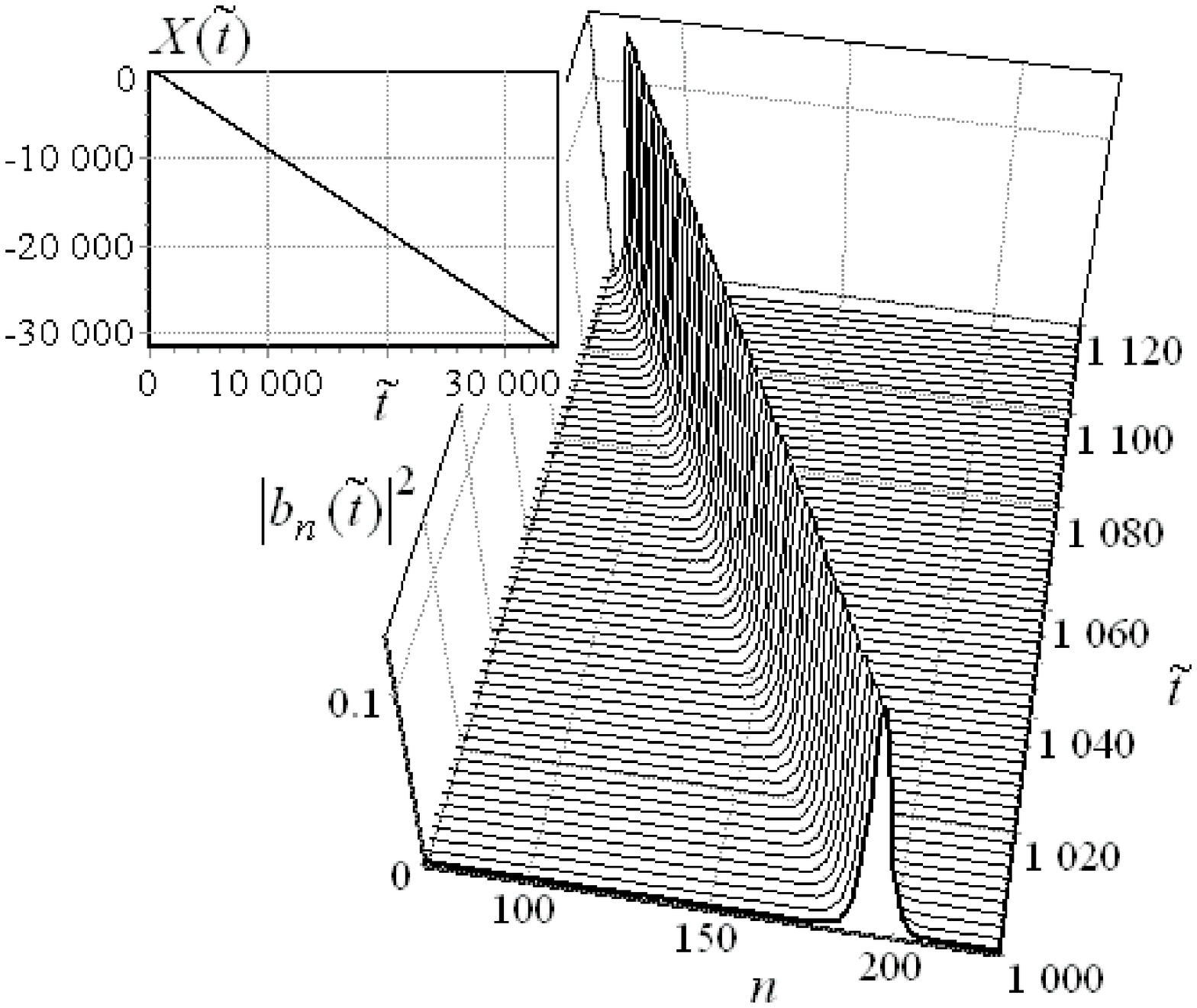}}(a)\\
\resizebox{0.43\textwidth}{!}{\includegraphics{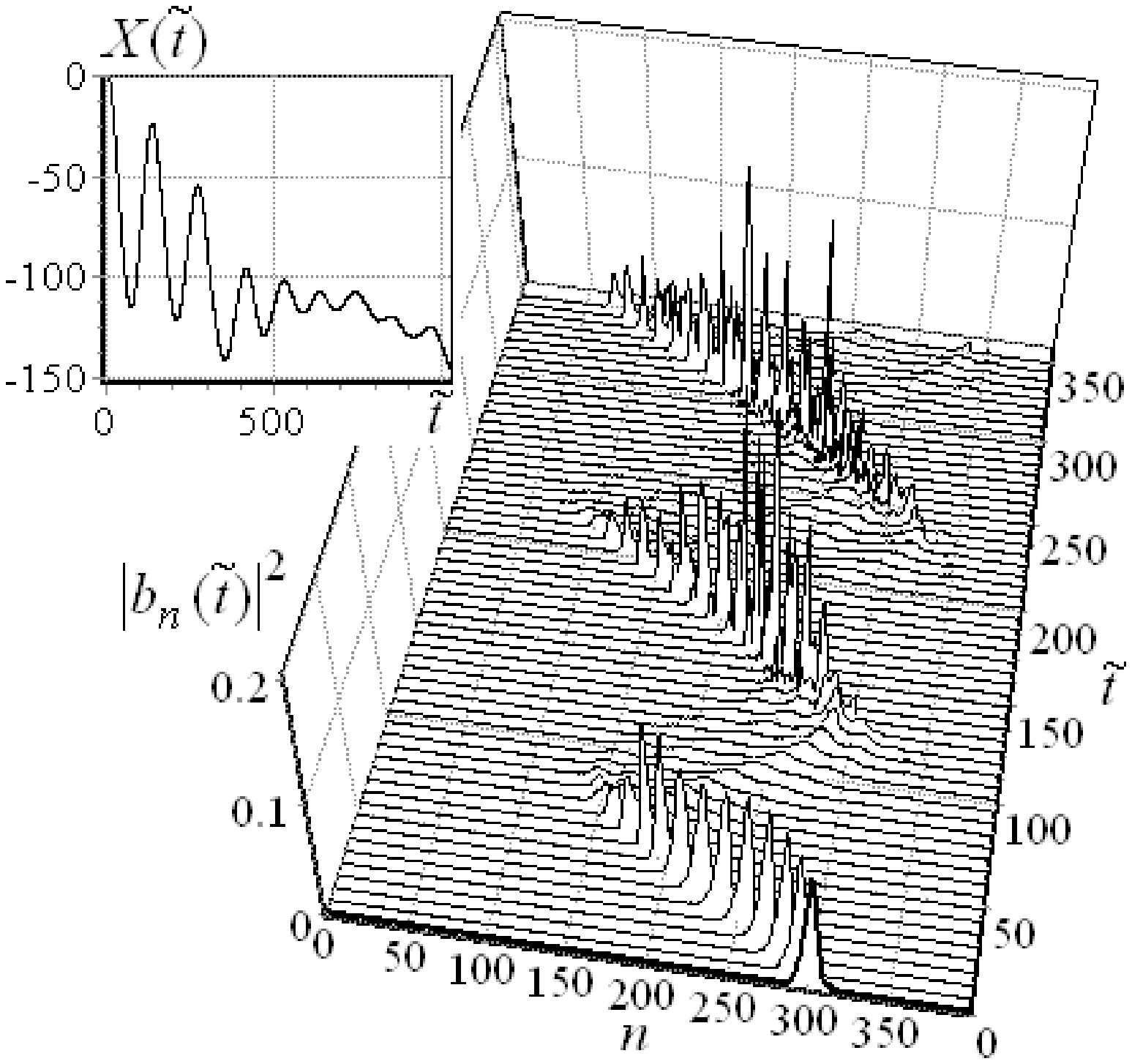}}(b)
\caption{Dependence of the charge motion character on the electric field intensity $E$ for $\kappa=1, \omega=1, \eta=1.276$. When $E=0.0005$ (fig.(a)) the polaron moves at a constant velocity, for $E=0.05$ (fig.(b)), the initial polaron motion at once turns to Bloch oscillations and then its motion takes the form of a breathing mode of Bloch oscillations.
}\label{fig_4}
\end{center}
\end{figure}
At small values of the electric field intensity, when the polaron velocity is small, its motion is not accompanied by Bloch oscillations Figure~\ref{fig_4}a. The reason is that the energy of excited oscillators in the "tail" of the moving polaron increases during the period of Bloch oscillations and exceeds greatly their energy. It can be said that while moving along the chain a polaron emits phonons. In a conservative system, such emission plays the role of friction, prohibiting the polaron from accelerating in an electric field. This makes possible polaron's motion at a constant velocity in a weak electric field.

As the electric field intensity increases, the situation changes: the energy of Bloch oscillations grows, while the length of the "tail" which has grown in the course of oscillations, decreases. When the energy of Bloch oscillations exceeds the "tail's"\, energy, they become visible Figure~\ref{fig_4}b. Depending on the values of the system's parameters Bloch oscillations can look like oscillations of a polaron as a whole, or may take the form of a breathing mode of Bloch oscillations.

The period of Bloch oscillations in dimensionless variables is $\widetilde{T}_{BO}=2\pi\big/E$ ($T_{BO}=2\pi\hbar\big/\tau\mathcal{E}ea$, $E=\mathcal{E}ea\tau\big/\hbar$). If the electric field intensity $E=0.05$ the period of Bloch oscillations is  $\widetilde{T}_{BO}\approx125.66$. Figure~\ref{fig_4}b demonstrates some periods of Bloch oscillations at the onset of the polaron motion in the electric field of intensity $E=0.05$. The graph of the function $|b_n(\tilde{t})|^2$ on Figure \ref{fig_4}a demonstrates a small segment of the polaron movement at a constant speed in the electric field of intensity $E=0.0005$. Figure~\ref{fig_4}a also represents the graph of the function $X(\widetilde{t})$ (\ref{11}) for the time $\widetilde{t}=34400$. The length of a chain for this case was chosen to be $N=32000$. The period of Bloch oscillations for the electric field intensity $E=0.0005$ is $\widetilde{T}_{BO}\approx12566$. Hence we observe a stationary motion of a polaron (without any oscillations) for the time corresponding to almost three periods of Bloch oscillations for $E=0.0005$. At the end of calculations ($\widetilde{t}=34400$) we observe the same graph of the function $|b_n (\tilde{t})|^2$, as presented in Figure~\ref{fig_4}a ($\widetilde{t}=1000$). These results are in a good agreement with theoretical conclusions  \cite{12}, Figure~\ref{fig_3}b.

In order to estimate numerically the critical value of the electric field $E_{cr}$ at which Bloch oscillations ($BO$) arise let us set the energy of oscillators on the interval which the polaron runs during the period of Bloch oscillations: $W_{osc}=\hbar\omega n_{osc}$ (where $n_{osc}=vT_{BO}\big/a$ is the number of oscillating sites on the interval $vT_{BO}$, $T_{BO}$ is the period of Bloch oscillations) equal to the energy of Bloch oscillations $W_{BO}=\hbar\omega_{BO}, \hbar\omega_{BO}=e\mathcal{E}a$. As a result, the field's critical value will be $E_{cr}=\sqrt{2\pi\omega V}$. The electric field is weak: $E<E_{cr}$, if the energy of Bloch oscillations is less than the loss of energy by polaron (caused by excitation of oscillations in the chain while the polaron moves along it) during the period of the polaron's $BO$. Accordingly, the field is strong if $E>E_{cr}$.

The upper branch of the curve for the $V(E)$ dependence in Figure~\ref{fig_3}a  is unstable and is not reproduced in experiments. When $E>E_{max}$, where $E_{max}$ is the maximum electric field at which the steady motion of the polaron is possible, (in numerical experiments this happens well before $E=E_{max}$ is achieved) Bloch oscillations become dominating. In this case the energy which the electron gets from the electric field goes into Bloch oscillations which, in turn, transfer it to the oscillators of the chain.

The velocity at which the mass center of a Bloch oscillator moves in the electric field is also limited and reaches its maximum at $E\leq E_{max}$. In the limit $E\rightarrow\infty$ this velocity vanishes, since the $BO$ amplitude turns into zero and the polaron localizes at one site.

The results obtained testify to the complicated character of the charge motion in a molecular chain.

The statement that a Holstein polaron cannot move in a molecular chain with dispersion-free oscillations of sites was first made by Davydov and Enol'skii in \cite{19}. In their subsequent works this statement was generalized to the case of polar crystals with optical phonons \cite{20}, \cite{21}. Some erroneous works devoted to this problem were analyzed in \cite{12}.

In conclusion it may be said that in this work we have carried out a direct numerical modeling of the Holstein polaron motion in a dispersion-free chain. The modeling demonstrated that in an electric field $E<E_{cr}$ the polaron moves along the chain at a constant velocity. In the range of parameter values at which the polaron size becomes comparable with the lattice constant, its motion is accompanied by oscillations of its shape (height and width of distribution $|b_n|^2$). In the course of oscillations the polaron velocity $V$ remains invariant. In the conservative system under consideration this invariance is caused by the fact that while traveling along the chain, the polaron excites oscillations of sites.

When $E>E_{cr}$, the steady motion of the polaron becomes impossible: the energy obtained by the polaron from the field per unit time becomes greater than that lost to the excitation of oscillations in the chain. The steady motion in this case "falls" and turns to Bloch oscillations. As distinct from Bloch oscillations in a non-deformable chain when the oscillation center is left in place (finite motion), in a deformable chain the polaron's mass center demonstrates translational movement accompanied by Bloch oscillations (infinite motion). In a strong field the polaron motion becomes similar to the breathing mode of Bloch oscillations.

Though in this work we have considered the case of a one-dimensional Holstein chain, the results obtained are rather general and can be applied to any media with dispersion-free phonons in which formation of a localized polaron state is possible. In particular, these results are valid for a strong coupling polaron in ionic crystals.

\begin{acknowledgement}
    The work was done with the support from the RFBR, projects
10-07-00112-a; 09-07-12073-ofi-m.\\
The authors are thankful to the Joint Supercomputer Center of the
Russian Academy of Sciences, Moscow, Russia for the provided
computational resources.\\
\end{acknowledgement}

\end{document}